\documentclass{emulateapj}

\usepackage{amssymb,amsmath,natbib,threeparttable,rotating}
\usepackage[graphicx]{realboxes}
\usepackage{ulem,color}

\begin{document}

\title{A substantial population of massive quiescent galaxies at
$\mathrm{z\sim4}$ from ZFOURGE\altaffilmark{12}}
\author{Caroline M. S. Straatman\altaffilmark{1}, Ivo Labb\'e\altaffilmark{1}, Lee R. Spitler\altaffilmark{2,11}, Rebecca Allen\altaffilmark{3,11}, Bruno Altieri\altaffilmark{4}, Gabriel B. Brammer\altaffilmark{5}, Mark Dickinson\altaffilmark{6}, Pieter van Dokkum\altaffilmark{7}, Hanae Inami\altaffilmark{6}, Karl Glazebrook\altaffilmark{3}, Glenn G. Kacprzak\altaffilmark{3,10}, Lalit Kawinwanichakij\altaffilmark{8}, Daniel D. Kelson\altaffilmark{9}, Patrick J. McCarthy\altaffilmark{9}, Nicola Mehrtens\altaffilmark{8}, Andy Monson\altaffilmark{9}, David Murphy\altaffilmark{9}, Casey Papovich\altaffilmark{8}, S. Eric Persson\altaffilmark{9}, Ryan Quadri\altaffilmark{9}, Glen Rees\altaffilmark{2}, Adam Tomczak\altaffilmark{8}, Kim-Vy H. Tran\altaffilmark{8}, Vithal Tilvi\altaffilmark{8}}

\altaffiltext{1}{Leiden Observatory, Leiden University, PO Box 9513, 2300 RA Leiden, The Netherlands; straatman@strw.leidenuniv.nl}
\altaffiltext{2}{Department of Physics and Astronomy, Faculty of Sciences, Macquarie University, Sydney, NSW 2109, Australia}
\altaffiltext{3}{Centre for Astrophysics and Supercomputing, Swinburne University, Hawthorn, VIC 3122, Australia}
\altaffiltext{4}{European Space Astronomy Centre (ESAC)/ESA, Villanueva de la Ca\~nada, 28691, Madrid, Spain}
\altaffiltext{5}{European Southern Observatory, Alonso de C\'ordova 3107, Casilla 19001, Vitacura, Santiago, Chile}
\altaffiltext{6}{National Optical Astronomy Observatory, Tucson, AZ, USA}	
\altaffiltext{7}{Department of Astronomy, Yale University, New Haven, CT 06520, USA}
\altaffiltext{8}{George P. and Cynthia W. Mitchell Institute for Fundamental Physics and Astronomy, Department of Physics and Astronomy, Texas A\& M University, College Station, TX 77843}
\altaffiltext{9}{Carnegie Observatories, Pasadena, CA 91101, USA}
\altaffiltext{10}{Australian Research Council Super Science Fellow}	
\altaffiltext{11}{Australian Astronomical Observatories, PO Box 915, North Ryde NSW 1670, Australia}
\altaffiltext{12}{This paper contains data gathered with the 6.5 meter Magellan Telescopes located at Las Campanas observatory, Chile.}

\begin{abstract}
We report the likely identification of a substantial population of massive $M\sim10^{11}M_{\odot}$ galaxies at $z\sim4$ with suppressed star formation rates (SFRs), selected on rest-frame optical to near-IR colors from the FourStar Galaxy Evolution Survey. The observed spectral energy distributions show pronounced breaks, sampled by a set of near-IR medium-bandwidth filters, resulting in tightly constrained photometric redshifts. Fitting stellar population models suggests large Balmer/$\mathrm{4000\AA}$ breaks, relatively old stellar populations, large stellar masses and low SFRs, with a median specific SFR of $2.9\pm1.8\times10^{-11}$/yr. Ultradeep Herschel/PACS $100\micron$, $160\micron$ and Spitzer/MIPS $24\micron$ data reveal no dust-obscured SFR activity for 15/19(79\%) galaxies. Two far-IR detected galaxies are obscured QSOs. Stacking the  far-IR undetected galaxies yields no detection, consistent with the SED fit, indicating independently that the average specific SFR is at least $10\times$ smaller than of typical star-forming galaxies at $z\sim4$. Assuming all far-IR undetected galaxies are indeed quiescent, the volume density is $\mathrm{1.8\pm0.7\times10^{-5}Mpc^{-3}}$ to a limit of $\mathrm{log_{10}}M/M_{\odot}\geq10.6$, which is $10\times$ and $80\times$ lower than at $z=2$ and $z=0.1$. They comprise a remarkably high fraction($\sim35$\%) of $z\sim4$ massive galaxies, suggesting that suppression of star formation was efficient even at very high redshift. Given the average stellar age of $0.8$Gyr and stellar mass of $0.8\times10^{11} M_{\odot}$, the galaxies likely started forming stars before $z=5$, with SFRs well in excess of $100M_\sun$/yr, far exceeding that of similarly abundant UV-bright galaxies at $z\geqslant4$. This suggests that most of the star formation in the progenitors of quiescent $z\sim4$ galaxies was obscured by dust.
\end{abstract}

\section{Introduction}

\begin{figure*}
\begin{center}
\includegraphics[width=0.7\textwidth]{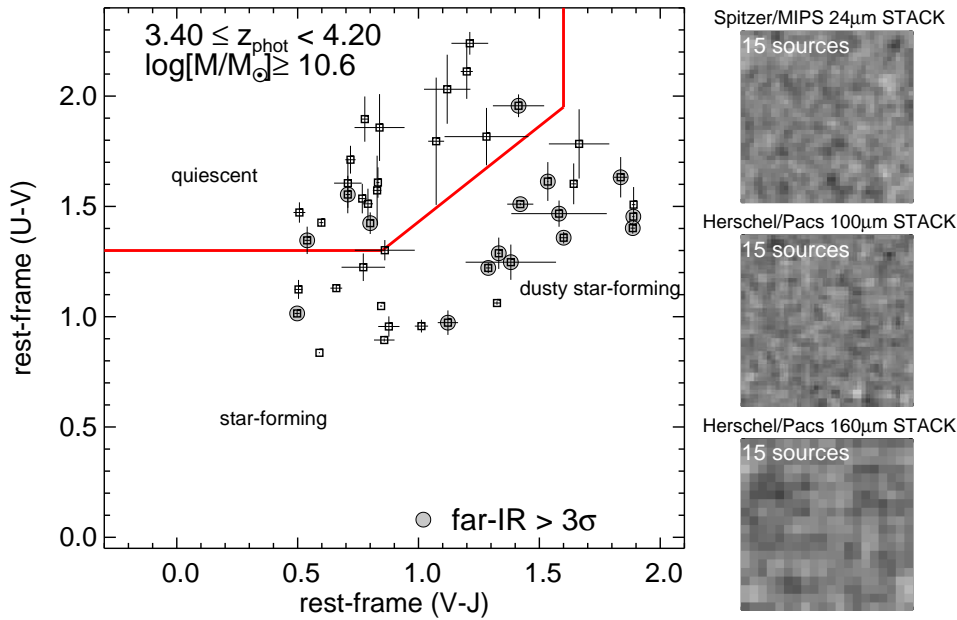}
\caption{Left: Rest-frame UVJ diagram of galaxies with $\mathrm{log_{10}}M/M_{\odot}\geq10.6$ at $\mathrm{3.4\leq\mathit{z}<4.2}$. 
The red solid line separates quiescent galaxies (top-left region) from star-forming galaxies. Galaxies with $\geq3\sigma$ far-IR detections are indicated with gray filled symbols and tend to be found amongst dusty star-forming galaxies. 19 objects are classified as quiescent, with 15/19 far-IR undetected.
Right: $\mathrm{24\micron}$, $\mathrm{100\micron}$ and $\mathrm{160\micron}$ stacks ($48\arcsec\times48\arcsec$) of undetected quiescent galaxies. Stacking yields no detection: $S_{24\micron}<0.002$mJy, $S_{100\micron}<0.090$mJy and $S_{160\micron}<0.140$mJy ($1\sigma$).} 
\label{fig:uvj}
\end{center}
\end{figure*}

The identification of a population of compact
quiescent galaxies at $1<z<3$, characterized by suppressed star formation and very small sizes, has attracted a significant interest \citep[e.g.][]{Daddi05,Dokkum08}. 
It remains an open question when these galaxies first appeared. The declining number densities and fractions of quiescent galaxies with redshift at $1<z<3$ suggest they might be rare at $z>3$ \cite[e.g.][]{Muzzin13b}. Nevertheless, the high ages of some quiescent galaxies at $z\sim2$ \citep{Whitaker13} suggest they could have
already existed at much earlier times. If confirmed at $z>3$, massive galaxies must have formed rapidly, early and with an effective mechanism of suppressing star formation.

Beyond $z=3$ candidate early-type or post-starburst galaxies have been reported, despite uncertainties whether their red colors could be due to dust-reddening \citep[e.g.][]{Chen04,Wiklind08,Mancini09,Fontana09,Guo13,Stefanon13,Muzzin13b}. In this Letter, we use the FourStar Galaxy Evolution Survey\footnote{http://zfourge.tamu.edu}  (ZFOURGE; Labb\'e et al. in preparation) to look for the earliest examples of quiescent galaxies. The strength of ZFOURGE lies in the unique combination of depth and the medium-bandwidth filters (covering $1-2\micron$) of the FourStar Infrared Camera \citep{Persson13} on the 6.5m Magellan Baade Telescope. 
These enable the derivation of accurate photometric redshifts and the detection of the age-sensitive Balmer/$\mathrm{4000\AA}$ break in faint, red galaxies at $1<z<4.2$. 

Throughout, we assume a $\mathrm{\Lambda CDM}$ cosmology with $\mathrm{\Omega_M=0.3,\Omega_{\Lambda}=0.7}$ and $H_0=70\mathrm{kms^{-1} Mpc^{-1}}$. The photometric system is AB.

\section{Data}\label{sec:data}

ZFOURGE covers three $\mathrm{11'\times11'}$ pointings in the fields CDFS, COSMOS and UDS, to very deep limits ($\sim26$ AB total mag ($5\sigma$) in $J_{1},J_{2},J_{3}$ and $\sim25$ mag in $H_s,H_l$ and $K_s$).  We combine ZFOURGE with public data, including HUGS (PI:Fontana) HAWK-I and CANDELS \citep{Grogin11,Koekemoer11} HST/WFC3 imaging, over a wavelength range of 0.3-8 $\mu$m. Full photometric $K_s-$band selected catalogs will be presented in Straatman et al. (in preparation).

We use Spitzer/MIPS $\mathrm{24\micron}$ data from GOODS-South (PI: Dickinson), COSMOS (PI: Scoville) an	d SPUDS (PI: Dunlop) and ultradeep Herschel/PACS $\mathrm{100\micron}$ and $\mathrm{160\micron}$ imaging from the GOODS-Herschel \citep{Elbaz11} and the CANDELS-Herschel program (PI:Dickinson), to independently place constraints on the on-going SFR. The ultradeep PACS $160\micron$ imaging currently provides the best sensitivity for far-IR light from star formation at high redshift, trading off k-correction and source confusion due to increasing beam size \citep{Elbaz11}, while $24\micron$ data are more sensitive to the presence of hot dust associated with AGN.

Photometric redshifts and rest-frame colors were derived with EAZY
\citep{Brammer08}. Comparing ZFOURGE photometric redshifts to spectroscopic redshifts, \cite{Tomczak13} found a scatter of $\sigma_{{\delta}z/(1+z)}=0.019$. Stellar population properties were derived by fitting \cite{Bruzual03} models with FAST \citep{Kriek09}, assuming a \cite{Chabrier03} initial mass function, exponentially declining star formation histories with timescale $\tau$, and solar metallicity.

\begin{table*}
\caption{Quiescent galaxies at $\mathrm{3.4\leq\mathit{z}<4.2}$}
\Rotatebox{90}{\begin{threeparttable}
\begin{tiny}
\begin{tabular}{l r r r r r r r r r r r r r r r}
\hline
\hline
& ra & dec & & & & M & $\mathrm{SFR_{SED}}$ & $\mathrm{sSFR_{SED}}$ & $\tau$\tnote{b} & age & & Ks\_tot\tnote{c} & $\mathrm{24\mu m}$\tnote{d} & $\mathrm{100\mu m}$\tnote{d} & $\mathrm{160\mu m}$\tnote{d} \\
ID & (deg) & (deg) & $\mathrm{z_{phot}}$ & U-V & V-J & $\mathrm{(10^{11}M_{\odot})}$ & $\mathrm{(M_{\odot}/yr)}$ & $\mathrm{(10^{-11}/yr)}$ & (Gyr) & (Gyr) & A(V) & (AB) & $\mathrm{(mJy)}$ & $\mathrm{(mJy)}$ & $\mathrm{(mJy)}$ \\
\hline
ZF-CDFS-209 &  53.1132774 & $-$27.8698730 & 3.56$ \pm $0.05 & 1.43$ \pm $0.02 & 
0.60$ \pm $0.01 &  0.76 &  2.239 &  2.884 & 0.10 & 0.63 & 0.3 & 22.6 & 
$-$0.001$ \pm$0.004 &  0.163$ \pm$0.163 & $-$0.047$ \pm$0.123\\
ZF-CDFS-403 &  53.0784111 & $-$27.8598385 & 3.660\tnote{a} & 1.42$ \pm $0.05 & 
0.80$ \pm $0.00 &  1.15 & 31.623 & 27.542 & 0.25 & 0.79 & 0.8 & 22.4 & 
 0.100$\pm$0.005\tnote{*$\dagger\times$} & 
 1.272$ \pm$0.199\tnote{*$\dagger\times$} & 
 1.686$ \pm$0.201\tnote{*$\dagger\times$}\\
ZF-CDFS-617 &  53.1243553 & $-$27.8516121 & 3.700\tnote{a} & 1.35$ \pm $0.06 & 
0.54$ \pm $0.01 &  0.69 & 13.183 & 18.621 & 0.16 & 0.63 & 0.3 & 22.3 & 
 0.087$\pm$0.003\tnote{*$\dagger$} &  1.062$ \pm$0.152\tnote{*$\dagger$} & 
 0.362$ \pm$0.157\tnote{$\dagger$}\\
ZF-CDFS-4719 &  53.1969414 & $-$27.7604313 & 3.59$ \pm $0.14 & 1.54$ \pm $0.07
 & 0.77$ \pm $0.02 &  0.45 &  0.851 &  1.905 & 0.16 & 1.00 & 0.3 & 23.4 & 
$-$0.000$ \pm$0.004 & $-$0.339$ \pm$0.175 & $-$0.287$ \pm$0.146\\
ZF-CDFS-4907 &  53.1812820 & $-$27.7564163 & 3.46$ \pm $0.16 & 1.57$ \pm $0.16
 & 0.83$ \pm $0.02 &  0.40 &  0.000 &  0.000 & 0.01 & 0.40 & 0.8 & 23.6 & 
 0.001$\pm$0.004 &  0.302$ \pm$0.132 &  0.154$ \pm$0.107\\
ZF-CDFS-5657 &  53.0106506 & $-$27.7416019 & 3.56$ \pm $0.07 & 1.61$ \pm $0.07
 & 0.83$ \pm $0.01 &  0.76 &  3.311 &  4.467 & 0.25 & 1.26 & 0.3 & 23.0 & 
 0.001$\pm$0.005\tnote{$\dagger$} &  0.193$ \pm$0.259\tnote{$\dagger$} & 
 0.078$ \pm$0.214\tnote{$\dagger$}\\
ZF-COSMOS-13129 & 150.1125641 &  2.3765368 & 3.81$ \pm $0.17 & 1.96$ \pm $0.05
 & 1.41$ \pm $0.11 &  1.78 &  0.000 &  0.000 & 0.01 & 1.58 & 0.6 & 23.6 & 
 0.112$\pm$0.010\tnote{*} &  0.895$ \pm$0.356 & $-$0.215$ \pm$0.343\\
ZF-COSMOS-13172 & 150.0615082 &  2.3786869 & 3.55$ \pm $0.06 & 1.90$ \pm $0.10
 & 0.78$ \pm $0.01 &  1.45 &  0.000 &  0.000 & 0.04 & 0.79 & 0.6 & 22.4 & 
 0.004$\pm$0.007 & $-$0.007$ \pm$0.394 & $-$0.323$ \pm$0.420\\
ZF-COSMOS-13414 & 150.0667114 &  2.3823516 & 3.57$ \pm $0.19 & 1.61$ \pm $0.11
 & 0.71$ \pm $0.06 &  0.44 &  0.035 &  0.079 & 0.10 & 1.00 & 0.2 & 23.4 & 
 0.009$\pm$0.008 & $-$0.330$ \pm$0.546 &  0.194$ \pm$0.452\\
ZF-UDS-885 &  34.3685074 & $-$5.2994704 & 3.99$ \pm $0.41 & 1.80$ \pm $0.29 & 
1.07$ \pm $0.03 &  0.60 &  0.001 &  0.001 & 0.03 & 0.40 & 1.3 & 24.0 & 
 0.012$\pm$0.008 &  1.009$ \pm$0.537 & $-$0.089$ \pm$0.400\\
ZF-UDS-1236 &  34.3448868 & $-$5.2925615 & 3.58$ \pm $0.08 & 1.47$ \pm $0.05 & 
0.51$ \pm $0.02 &  0.60 &  0.550 &  0.912 & 0.06 & 0.50 & 0.4 & 22.6 & 
$-$0.016$ \pm$0.011 &  0.383$ \pm$0.449 &  0.537$ \pm$0.360\\
ZF-UDS-2622 &  34.2894516 & $-$5.2698011 & 3.77$ \pm $0.10 & 1.51$ \pm $0.07 & 
0.79$ \pm $0.02 &  0.87 & 16.218 & 18.621 & 0.16 & 0.63 & 0.9 & 23.0 & 
 0.013$\pm$0.010 &  0.761$ \pm$0.372 &  0.152$ \pm$0.442\\
ZF-UDS-3112 &  34.2904282 & $-$5.2620673 & 3.53$ \pm $0.06 & 1.71$ \pm $0.06 & 
0.72$ \pm $0.02 &  0.43 &  1.862 &  4.467 & 0.25 & 1.26 & 0.0 & 23.2 & 
$-$0.010$ \pm$0.010 & $-$0.547$ \pm$0.377 &  0.256$ \pm$0.333\\
ZF-UDS-5418 &  34.2937546 & $-$5.2269468 & 3.53$ \pm $0.07 & 1.55$ \pm $0.08 & 
0.71$ \pm $0.01 &  0.44 &  3.020 &  6.761 & 0.16 & 0.79 & 0.5 & 23.3 & 
 0.049$\pm$0.010\tnote{*} &  0.560$ \pm$0.443 & $-$0.494$ \pm$0.386\\
ZF-UDS-6119 &  34.2805405 & $-$5.2171388 & 4.05$ \pm $0.27 & 1.86$ \pm $0.15 & 
0.84$ \pm $0.10 &  0.55 &  5.623 & 10.471 & 0.10 & 0.50 & 1.0 & 23.8 & 
$-$0.013$ \pm$0.009 &  0.224$ \pm$0.477 &  0.331$ \pm$0.316\\
ZF-UDS-9526 &  34.3381844 & $-$5.1661916 & 3.97$ \pm $0.18 & 2.11$ \pm $0.12 & 
1.20$ \pm $0.03 &  0.89 & 16.596 & 18.621 & 0.16 & 0.63 & 1.8 & 24.2 & 
 0.016$\pm$0.009 &  0.038$ \pm$0.351 & $-$0.000$ \pm$0.294\\
ZF-UDS-10401 &  34.3601379 & $-$5.1530914 & 3.91$ \pm $0.38 & 1.82$ \pm $0.13 & 
1.28$ \pm $0.17 &  0.38 &  0.001 &  0.002 & 0.02 & 0.25 & 1.7 & 24.6 & 
 0.007$\pm$0.010 & $-$0.500$ \pm$0.340 & $-$0.490$ \pm$0.383\\
ZF-UDS-10684 &  34.3650742 & $-$5.1488328 & 3.95$ \pm $0.48 & 2.03$ \pm $0.16 & 
1.12$ \pm $0.10 &  0.85 &  3.802 &  4.467 & 0.25 & 1.26 & 1.0 & 24.1 & 
 0.007$\pm$0.012 & $-$0.105$ \pm$0.388 & $-$0.503$ \pm$0.544\\
ZF-UDS-11483 &  34.3996315 & $-$5.1363320 & 3.63$ \pm $0.32 & 2.24$ \pm $0.05 & 
1.21$ \pm $0.08 &  1.02 &  0.000 &  0.000 & 0.01 & 1.00 & 1.0 & 23.6 & 
 0.004$\pm$0.011 & $-$0.149$ \pm$0.409 &  0.715$ \pm$0.325\\
\hline
\end{tabular}
\end{tiny}
\begin{tablenotes}
\item[a]Spectroscopic redshift from \cite{Szokoly04}.
\item[b]$\mathrm{SFR\sim exp(-t/\tau)}$.
\item[c]Total magnitude in the FourStar $K_s-$band.	
\item[d]``*": $\geq3\mathrm{\sigma}$ detections. ``$^{\dagger}$": X-ray detected \citep{Szokoly04,Xue11}. ``$^{\times}$": radio source \citep{Miller13}.
\end{tablenotes}
\end{threeparttable}}	
\label{tab:1}
\end{table*}

\begin{figure*}
\includegraphics[width=\textwidth]{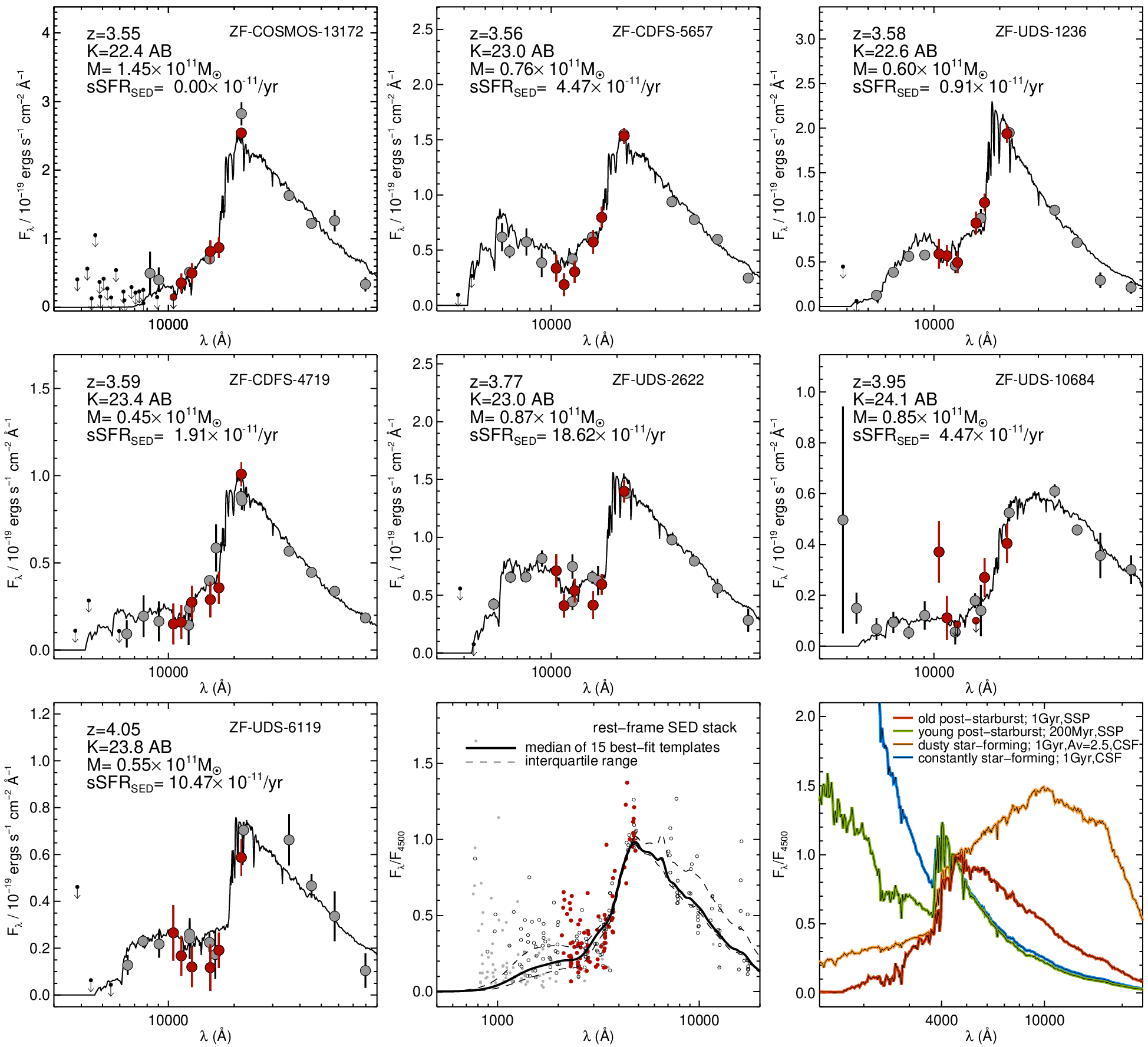}
\caption{Observed SEDs of UVJ selected quiescent galaxies. Red datapoints correspond to the FourStar medium-bandwidth
filters. The solid curve is the fitted model from FAST.
Downward pointing arrows are $1\sigma$ upper limits. 
Bottom-middle: Rest-frame SED of the 15 far-IR undetected galaxies (open symbols), normalized at rest-frame $4500$\AA, with gray symbols corresponding to $1\sigma$ upper limits. The solid curve is the median of the best-fit template SEDs.  Dashed lines mark the interquartile range. Bottom-right: Four model SEDs with constant star formation or a single stellar population (SSP) and ages from 200Myr to 1Gyr. The observed SEDs are characterized by pronounced Balmer/$\mathrm{4000\AA}$ breaks, similar to the old post-starburst model.}
\label{fig:sed}
\end{figure*}

\section{Selection of quiescent galaxies at $z\sim 4$.}\label{sec:uvj}

We use a two-color criterion (rest-frame $U-V$ versus $V-J$; Figure \ref{fig:uvj}) to separate quiescent galaxies (red in $U-V$, but blue in $V-J$) from star-forming galaxies, (blue or red in both $U-V$ and $V-J$ colors) \citep[e.g.,][]{Labbe05,Williams09}. This technique has been shown to isolate the red sequence of galaxies at $z<3$ \citep[e.g.,][]{Whitaker11} and was spectroscopically confirmed to identify quiescent galaxies at $z\sim2$ \citep[e.g.][]{Whitaker13}. 

We focus on the redshift range $3.4\leq\mathit{z}<4.2$, where the medium-bandwidth filters straddle the Balmer/$\mathrm{4000\AA}$ break. At $z>3.4$ the break enters the $H_l$ filter ($1.7\micron$), while at $z<4.2$ the $K_s-$band ($2.2\micron$) still probes light redward
of the break. We limit the sample to a signal-to-noise of $>$7 in $K_s$ and stellar masses of $\mathrm{log_{10}}M/M_{\odot}\geq10.6$, where we are complete for passively evolving stellar populations formed at $z<10$. This yields 44 galaxies with high quality photometry, of which 15 fall in the UVJ quiescent region and are undetected in the FIR, a significant fraction: $34\pm 13$\% (15/44). A summary of their properties is presented in Table \ref{tab:1}. Their photometric redshifts range from $z=3.46$ to 
$z=4.05$ with a mean of $z=3.7$ and mean uncertainty ${\delta}z/(1+z)=0.036$, leading to well constrained rest-frame colors. Two galaxies have spectroscopic redshifts, with a mean $(z_{phot}-z_{spec})/(1+z_{spec})=-0.039$ \citep{Szokoly04}.

\section{properties of quiescent galaxies at $z\sim4$}\label{sec:prop}

\subsection{Spectral energy distributions}

We show representative SEDs of seven galaxies in Figure \ref{fig:sed}. The median SED of all 15 (far-IR undetected) quiescent 
galaxies, constructed by de-redshifting their photometry and normalising them to the flux density at $\mathrm{4500\AA}$, is also shown. 

The observed SEDs are exceedingly faint in the optical ($I\sim27$ magnitude)
and extremely red throughout the near-IR (median $I-Ks=3.7\pm0.33$). The SEDs are characterized by a sharp break, with $H-Ks=1.9\pm0.20$ and peaking in $K_s$, and a blue spectral slope in the mid-IR Spitzer/IRAC bands ($Ks-[4.5\micron]=0.91\pm0.13$). The break is reminiscent 
of the strong break found in quiescent galaxies at lower redshift, where it is
caused by combination of the Balmer and $\mathrm{4000\AA}$ absorption features, indicative of a combination of relatively old stellar populations and suppressed star formation. Additionally, some galaxies exhibit a  second break at bluer wavelengths, which is likely the Lyman break.

As it is difficult to separate the contribution of the Balmer break and the $\mathrm{4000\AA}$ break from photometry alone, we quantify the size of the total combined break ($D_{tot}$) by estimating the flux ($F_{\nu}$) ratio at $4000-4100\mathrm{\AA}$ and $3500-3650\mathrm{\AA}$ on the best-fit models {of each galaxy individually, extending the definition of D4000n \citep{Balogh99} to cover the Balmer break as well. We find the median $D_{tot} =2.8\pm0.1$, which is in the range of post-starburst galaxies with suppressed star formation (e.g., a $\tau=10$Myr model produces $D_{tot} =3.1$ at 500Myr). In contrast, unobscured constant star-forming (CSF) models only 
reach $D_{tot} =1.8$ at 1Gyr. Heavily obscured star-forming models (e.g. 1Gyr, CSF, $A_v=2.5$) 
can also reach quite red $D_{tot} =2.5$, but are ruled out as they predict very red $Ks-[4.5\micron]=2.0$, whereas the observed SED-slopes are bluer.

\begin{figure*}
\begin{center}
\includegraphics[width=\textwidth]{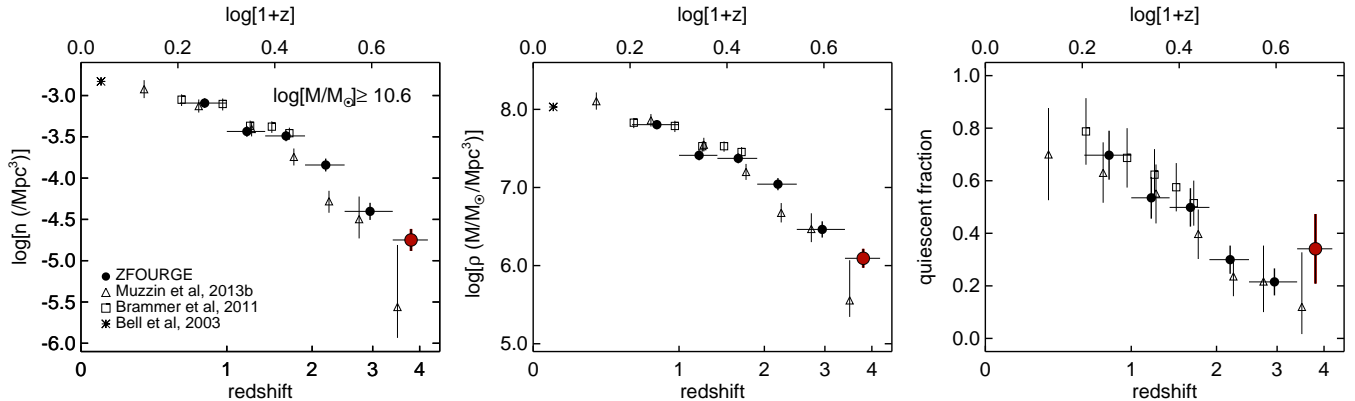}
\caption{Left: Number density of quiescent galaxies in ZFOURGE. Middle: Stellar mass density. Right: Quiescent fraction. Errors are a combination of the Poisson uncertainty and cosmic variance. Horizontal errorbars indicate the width of the redshift bins. The red symbols denote the 15 UVJ selected and far-IR undetected galaxies. We compare with \cite{Bell03} at $z=0.1$, \cite{Brammer11} to $z\lesssim2$ and \cite{Muzzin13b} at $0<z<4$.  The overall trend is a decrease in number density towards $z\sim4$, consistent with the earlier NMBS and UltraVISTA results. However, the larger depth and sampling of ZFOURGE allows for much better constraints on the evolution at $2<z<4$.  Surprisingly, at $z=3.7$, $34\pm13$\% of the galaxies with $\mathrm{log_{10}}M/M_{\odot}\geq10.6$ could be quiescent, suggesting that the decline of the quiescent fraction could flatten at $z\gtrsim2-3$.} 
\label{fig:nd}
\end{center}
\end{figure*}

\subsection{Stellar population fits}

Models with exponentially declining SFRs fit the data well, with a median $\chi^2_{red}=1.3$. The median best-fit age of the galaxies is $0.8$Gyr, the median star formation timescale ($\tau$) $0.1$Gyr, the average stellar mass $0.8\times10^{11}M_{\odot}$, and the median specific SFR (sSFR) $2.9\pm1.8\times10^{-11}$/yr.

To test if models with ongoing star formation provide acceptable fits to the sample, we force $\tau=250$Myr, $\tau=1$Gyr or CSF and refitted the data, finding a median $\chi^2_{red,250Myr}=2.1$, $\chi^2_{red,1Gyr}=6.6$ and $\chi^2_{red,CSF}=6.6$. This shows that models with $\tau=250$Myr provide almost equally good fits, but longer formation timescales ($\tau>1$Gyr) provide poor fits to the data. For all individual galaxies $\tau<250$Myr models produced better fits than did $\tau>1$Gyr models. We note that high redshift solutions with low sSFRs are preferred in all cases. Other solutions, e.g. at low redshift or with ongoing obscured star formation are ruled out at more than $99$\% confidence for 18/19 galaxies.

We refitted the data using the models of \cite{Maraston05}, and obtained a mean stellar mass of $0.5\times10^{11}M_{\odot}$, with a typical offset of $-0.2$ dex compared to the masses in Table \ref{tab:1}, and a median sSFR of $0.1\times10^{-11}$/yr. Hence the result is not strongly dependent on the adopted stellar population model.

Overall, the fits suggest most stars were formed at $z>5$, followed by an epoch of suppressed star formation. As expected, the median stellar ages are lower than the typical age of 1.3 Gyr found at $z\sim2$ by \cite{Whitaker13}. We find some galaxies with very red $U-V$  and $V-J$, pointing towards older stellar populations. However, their best-fit ages are the same as for the bluer galaxies, with larger redshift uncertainties or dust, suggesting that dust and photometric scatter are the main causes.

\subsection{Independent constraints on SFR and AGN activity from Herschel}\label{sec:fir}

We derived Spitzer/MIPS $24\micron$, Herschel/PACS $100\micron$ and $160\micron$ flux intensities, measured in apertures of $7\arcsec$, $8\arcsec$ and $12\arcsec$ diameter with aperture corrections of 2.56, 2.45 and 2.60, respectively (assuming a point source profile). Light from neighbouring sources was subtraced following \cite{Labbe10}. 

We find 1, 2 and 4 $>3\sigma$ detections at $160\micron$, $100\micron$ and $24\micron$, respectively. These 4/19 galaxies may have obscured star formation. The total detection rate ($21\pm 11$\%) is lower than the 50\% of $24\micron$ detections reported earlier for $z>3$ quiescent galaxies \citep{Stefanon13}. Two are also detected in X$-$ray and are type$-$2 QSOs \citep{Szokoly04,Xue11}, of which one is a radio source \citep{Miller13}. A third X$-$ray detection is found amongst the far-IR undetected galaxies, for a total of 3 likely AGN. 

To place tighter constraints on the average far-IR luminosity of the 15 far-IR undetected 
galaxies, we stack their $24\micron$, $100\micron$ and $160\micron$ images (Figure \ref{fig:uvj}), with uncertainties derived by bootstrap resampling. The formal measurements are $S_{24\micron}=0.001\pm0.002$mJy, $S_{100\micron}=0.049\pm0.090$mJy and $S_{160\micron}=0.039\pm0.140$mJy. Hence the sources are undetected. The strongest constraint on obscured SFR is obtained at $160\micron$. Using the IR templates of \cite{Wuyts11}, we find $7.1\pm25M_{\odot}$/yr. Given the mean mass of the sample ($0.8\times10^{11}M_{\odot}$), this corresponds to a sSFR of $0.9\pm3\times10^{-10}\mathrm{yr^{-1}}$. While these independently derived limits cannot rule out ongoing obscured star formation, they are consistent with the SED fits ($2.9\pm1.8\times10^{-11}$/yr), and are $\sim10\times$ smaller than the sSFR$=3-6\times10^{-9}$/yr of similarly massive galaxies at $z\sim3$ and typical UV-bright star-forming galaxies at $z\sim4$ \citep[e.g.][]{Stark13,Viero13}. 

\subsection{Contamination by emission lines}

We caution that the galaxies here could in fact be vigorously star-forming, if their $K_s-$band fluxes were boosted dramatically by emission lines ($[OIII]$ and $H\beta$), mimicking a Balmer Break \citep[e.g.][]{Shim11,Stark13}. We tested this scenario by fitting CSF models to the SEDs \textit{without fitting the $K_s-$band}, leading to fits with high obscuration (median $A(V)\sim2$).

The CSF models fit the data poorly and vastly underpredict the median $K_s-$band: $(K_{s,obs}-K_{s,SED})_{CSF}=-0.53\pm0.06$ (note that standard$-\tau$ models predict the $K_s-$band magnitude nearly perfectly: $(K_{s,obs}-K_{s,SED})_{\tau=free}=-0.01\pm0.05$). Assuming this excess is due to strong ($EW_{obs}\sim2000\mathrm{\AA}$) emission lines, the predicted median SFR is $\sim1000M_{\odot}$/yr, which, because of the high obscuration, should result in $7-18\sigma$ detections in $160\micron$, but is not observed. 

Furthermore, we can test the hypothesis that $K_s$ is boosted by $[OIII](\lambda\lambda4959,5007\mathrm{\AA})$ and $H\beta(4861\mathrm{\AA})$ at $3.0<z<3.6$, by looking at existing narrowband $NB209(2.10\micron)$ data, covering CDFS$+$COSMOS \citep{Lee12}. Since the odds are only 20\% that $NB209$ is affected by any of these lines, it effectively traces the continuum. Using a simple model, drawing uniformly random redshifts at $3.0<z<3.6$, the predicted median color is $(K_s-NB209)=-0.44$, nearly independent of line ratios, in strong disagreement with the observed median $(K_s-NB209)=-0.04\pm0.1$. We also inspected data from the 3D-HST survey \citep{Brammer12}, with low resolution spectral coverage at $1.1-1.6\micron$, as strong $[OIII]/H\beta$ lines in $K_s$ would imply strong [OII] in the HST/WFC3 grism. We found only 2 detections for 13 galaxies with coverage to an emission line sensitivity of $\sim5\times10^{-17}ergs/s/cm^{2}(5\sigma)$: $MgII(\lambda2798\mathrm{\AA})$ for the QSO ZF-CDFS-617 and $[OII]$ for the $24\micron$ detected ZF-UDS-5418.

\section{Implications}\label{sec:disc}

From hereon we adopt as operational definition of ``quiescent'': galaxies that satisfy the UVJ criterium and are not detected in the far-IR \citep[e.g.][]{Bell12}. We note however, that the current data do not allow to determine conclusively whether the galaxies have completely stopped forming stars as the sample is too faint for spectrographs on large telescopes.

\subsection{Number densities}

From the 15 quiescent galaxies we estimate the volume and stellar mass density, finding $\mathrm{1.8\pm 0.7\times 10^{-5}Mpc^{-3}}$ and $\mathrm{1.2\pm0.5\times10^6M_{\odot}Mpc^{-3}}$, respectively. Uncertainties are the quadratic sum of the Poission uncertainty and variations due to large scale structure \citep{Moster11}. The volume density in the $11\arcmin\times11\arcmin$ area in the ZFOURGE$-$UDS field is $\sim3\times$ higher than in the ZFOURGE$-$COSMOS field, underscoring the need for probing mulitple pointings to faint limits.

For comparison with other surveys, we integrated the COSMOS/UltraVISTA mass function at $3<z<4$ for quiescent galaxies of \cite{Muzzin13b}, based on a similar UVJ classification, to $\mathrm{log_{10}}M/M_{\odot}\geq10.6$. UltraVISTA produces a number density of $\mathrm{2.7\times10^{-6}Mpc^{-3}}$ and a mass density of $\mathrm{3.1\times 10^5M_{\odot}Mpc^{-3}}$. These are factors of $\sim 7$ and $\sim 4$, lower (albeit at only $\sim1.6\sigma$ significance). This is likely a completeness effect, as UltraVISTA is only complete to $M\gtrsim10^{11}M_{\odot}$. Indeed, \cite{Muzzin13b} select galaxies with $K_{s,tot,AB}<23.4$, while 50\% of the galaxies here have $K_{s,tot,AB}>23.4$. 

The number and stellar mass densities of quiescent galaxies at $0.6\leq\mathit{z}<4.2$ are shown in Figure \ref{fig:nd}. These were obtained from the full ZFOURGE catalogs (Straatman et al., in prep), using the same selection criteria as described in section \ref{sec:uvj}. The number density decreases rapidly towards $z\sim4$ ($\sim10\times$ lower than  at $z=2$ and $\sim80\times$ than at $z=0.1$), suggesting that a small fraction ($10-15\%$) of $z\sim2$ quiescent galaxies was already in place at $z\sim4$. The last panel of Figure \ref{fig:nd} shows the fraction of quiescent galaxies with $\mathrm{log_{10}}M/M_{\odot}\geq10.6$. This strongly declines with redshift between $0.6<z<3$. Therefore, we would expect a value close to zero at $z\sim4$, but we find a surprisingly high fraction of $34\pm13$\%. This is similar to the value at $z\sim2.2$ ($30\pm 8$\%), suggesting a flat quiescent fraction at $2<z<4$.

\subsection{Star-forming progenitors}

\begin{figure}
\includegraphics[width=0.49\textwidth]{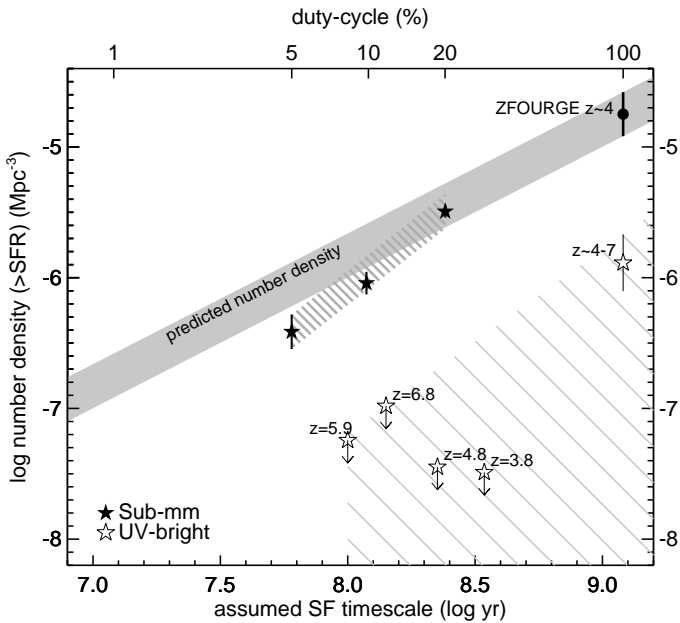}
\caption{The predicted cumulative number density ($\mathrm{n(>SFR)/Mpc^3}$) for the progenitors of the $z\sim4$ quiescent galaxies as a function of the assumed formation timescale. The gray filled area shows the expected range in number density based on the $z\sim4$ sample. If the galaxies form over a long timescale (e.g. $\sim1$ Gyr), this is simply the observed number density. If they form in shorter, more intense bursts (smaller duty-cycles), the observed progenitor number densities are expected to be smaller. Open star symbols are estimates from the SFR functions of \cite{Smit12} ($z\sim4-7$) and the UV-luminosity functions of \cite{Burg10} ($z=3.8$ and $z=4.8$), \cite{Willott13} ($z=5.9$) and \cite{Bowler12} ($z=6.8$). For the latter we assume that $\mathrm{log_{10}SFR}=-0.4(M_{1600}-\langle\mathit{A}_{1600}\rangle)-7.25$ \citep{Kennicutt98}, with $M_{1600}$ the luminosity at $1600\mathrm{\AA}$ in AB mag and $\langle A_{1600}\rangle$ the dust-correction factor from \cite{Bouwens12}. Upper limits are derived from the respective survey volumes. 
Filled stars show sub-mm number counts \citep{Karim13}, using a conversion of $1\mathrm{mJy}\approx1.667\times10^{12}L_{\odot,IR}$ \citep{Blain02} and $1M_{\odot}/yr=4.5\times10^{-44}L_{IR}(ergs/s)$ \citep{Kennicutt98} and assuming 10\% of these are at $z>4$.
The number of luminous UV-bright galaxies at $z>4$ is far too low (1-2 dex), while the sub-mm counts are much better matched, suggesting there might be sufficient numbers of heavily obscured starburst galaxies at $z>4$ if their formation timescales are $\sim$200Myr.}
\label{fig:pg}
\end{figure}

Given average stellar ages of $0.8$Gyr and masses of $0.8\times10^{11}M_{\odot}$, the galaxies likely started forming their stars much earlier than $z=5$, with SFRs well in excess of $100M_\sun$/yr. This raises the question what are the likely progenitors.
In recent years, UV-luminous galaxies have been 
found in large numbers to $z\sim10$ \citep[e.g.][]{Bouwens13,Ellis13}.
These are actively star-forming, although even the most luminous galaxies found so far at $z>4$ have relatively modest UV-derived SFRs ($<100M_\sun$/yr) \citep{Smit12}.  

The expected number density at $z>4$ of the progenitors depends on the assumed star formation timescale (gray shaded area in Figure \ref{fig:pg}). If the progenitors were visible at all times (i.e. a formation timescale of $\sim$1 Gyr and number density of $\mathrm{1.8\pm0.7\times10^{-5}Mpc^{-3}}$), then we can use the SFR functions at $z=4-7$ \citep{Smit12} to select progenitors with sufficiently high SFRs on fixed cumulative number density \citep{Dokkum10}.

As shown in Figure \ref{fig:pg}, this number density falls $\sim1.2$ dex short. If we assume shorter formation timescales (e.g. a few 100Myr), the progenitors require much higher SFRs and are predicted to be found in smaller numbers. Comparing to UV-luminosity functions from wide-field surveys, using the redshift window as the formation timescale, the number densities are $>1.5$ dex too low, reflecting that sufficiently luminous UV-bright galaxies are extremely rare. 

Alternatively, the main star formation episode is obscured by dust. There exists a population of high-redshift sub$-$mm detected galaxies, including highly obscured gas-rich mergers \citep[e.g.][]{Younger07}, with large SFRs ($\gtrsim1000M_\sun$/yr), that could be progenitors of $z\sim4$ quiescent galaxies. Based on the $870\micron$ source counts of \cite{Karim13}, and tentatively assuming that 10\% are at $z>4$ \citep[e.g.][]{Swinbank12}, we find that obscured starbursting galaxies are sufficiently numerous. This suggests that most of the star formation in the progenitors of quiescent $z\sim4$ galaxies could have been obscured by dust.

\section{Summary}\label{sec:sum}

Using very deep imaging from ZFOURGE we find evidence for the existence of massive ($M\sim10^{11}M_{\odot}$) galaxies with suppressed star formation at very early times ($z\sim4$). The galaxies satisfy the UVJ criterium, which has been shown to efficiently select quiescent galaxies at $z<3$ \citep{Whitaker11,Whitaker13,Williams09}. The observed SEDs show prominent breaks, well sampled by the FourStar near-IR medium-bands, leading to accurate photometric redshifts and illustrating a key strength of the survey. The SEDs are well fit by models with strong Balmer/$\mathrm{4000\AA}$ breaks, small $\tau (<250\mathrm{Myr})$, high ages ($\sim0.8$Gyr) and low sSFRs ($2.9\pm1.8\times10^{-11}$/yr). Consistent with this, 79\% of the galaxies are undetected in deep Spitzer/MIPS and Herschel/PACS imaging. Stacking the far-IR places an independent constraint on the average sSFR of $<3\times10^{-10}\mathrm{yr^{-1}}$, a factor $>10\times$ smaller than the average sSFR of UV-bright star-forming galaxies at these redshifts \citep[e.g.][]{Stark13} and consistent with these galaxies having strongly suppressed SFRs.

While rare (with number densities $\sim10\times$ and $\sim80\times$ lower than at $z=2$ and $z=0.1$), they make up a surprisingly high fraction of the massive galaxy population at $z\sim4$ ($34\pm13$\%), higher than expected based on the declining trend over $1<z<3$, suggesting an effective mechanism of suppressing star formation and short formation timescales ($<1$Gyr). The implied SFRs needed to form galaxies with a mean stellar mass of $0.8\times10^{11}M_{\odot}$ in such a short time exceeds that of similarly abundant UV-bright galaxies at $z\geqslant4$, suggesting that most of the star formation in their progenitors was obscured by dust.

We emphasize that without spectroscopic confirmation the number of quiescent galaxies at $z\sim4$ remains poorly constrained, but given their faint magnitudes, real progress will likely have to wait until the launch of JWST or construction of ELTs. Currently, ALMA observations can place stronger limits on the dust-obscured activity of these galaxies and help identify the progenitors at $z\gtrsim4$.

\section{Acknowledgements}
This research was supported in part by the George P. and Cynthia Woods Mitchell Institute for Fundamental Physics and Astronomy. We would especially like to thank the Mitchell family for their continuing support. We thank Danilo Marchesini and Adam Muzzin for help with the UltraVISTA number densities. We thank Marijn Franx, Mattia Fumagalli, Shannon Patel, Jesse van de Sande, David Sobral, Renske Smit, Paul van der Werf and Claudia Maraston for useful discussions. We acknowledge support by the following grants: NSF AST-1009707, ERC HIGHZ \#227749 and NL-NWO Spinoza. Australian access to the Magellan Telescopes was supported through the NCRIS of the Australian Federal Government.  This work is based on observations made with Herschel, an ESA Cornerstone Mission with significant participation by NASA, through an award issued by JPL/Caltech.

%\nocite{*}
\bibliographystyle{apj}

%\bibliography{apj-jour,./bibms}

\end{document}